\documentclass[aps,prl,twocolumn,showpacs,superscriptaddress,10pt]{revtex4-1}

\usepackage{amsmath}
\usepackage{amsfonts}
\usepackage{amssymb}
\usepackage{graphicx}
\usepackage{color}
\usepackage{xcolor}

\newcommand{\dd}{\mathrm{d}}
\newcommand{\ee}{\mathrm{e}}

\newcommand{\qq}{\mathbf{q}}

\usepackage[resetlabels]{multibib}
\newcites{sm}{SM sources}
\begin{document}


%
\title{Fluctuation-Induced Quantum Zeno Effect}
\author{Heinrich Fr\"{o}ml}
\affiliation{Institute for Theoretical Physics, University of Cologne, D-50937 Cologne, Germany}
\author{Alessio Chiocchetta}
\affiliation{Institute for Theoretical Physics, University of Cologne, D-50937 Cologne, Germany}
\author{Corinna Kollath}
\affiliation{HISKP, University of Bonn, 53115 Bonn, Germany}
\author{Sebastian Diehl}
\affiliation{Institute for Theoretical Physics, University of Cologne, D-50937 Cologne, Germany}
\affiliation{Kavli Institute for Theoretical Physics, University of California, Santa Barbara, CA 93106, USA}

\begin{abstract}
An isolated quantum gas with a localized loss features a non-monotonic behavior of the particle loss rate as an incarnation of the quantum Zeno effect, as recently shown in experiments with cold atomic gases. 
While this effect can be understood in terms of local, microscopic physics, we show that novel many-body effects emerge when non-linear gapless quantum fluctuations become important.  
To this end, we investigate the effect of a local dissipative impurity on a one-dimensional gas of interacting fermions. We show that the escape probability for modes close to the Fermi energy vanishes for an arbitrary strength of the dissipation. In addition, transport properties across the impurity are qualitatively modified, similarly to the Kane-Fisher barrier problem. We substantiate these findings using both a microscopic model of spinless fermions and a Luttinger liquid description.
\end{abstract}



\date{\today}
\maketitle

\begin{figure}[t!]
\centering
\includegraphics[width=8cm]{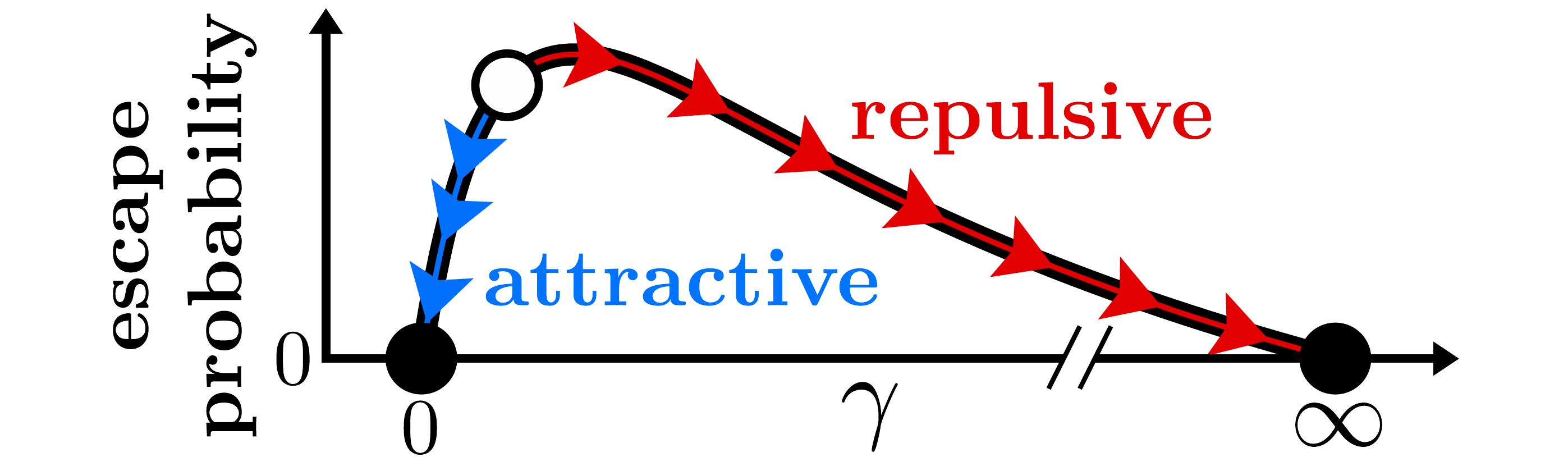}
\caption{(Color online). Non-monotonic behavior of the escape probability as a function of the dissipation strength. For momenta close to $k_F$, gapless fluctuations renormalize the escape probability, reaching the Zeno (rightmost black dot) or transparent (leftmost black dot) fixed points for repulsive or attractive interactions, respectively.}
\label{fig:fig1}
\end{figure}

\emph{Introduction ---} The quantum Zeno effect (QZE) entails that, perhaps surprisingly, a frequent measurement of a microscopic quantum system suppresses transitions between quantum states~\cite{Misra1977}. 
Recently, experiments with ultracold atoms have revealed the QZE in many-body systems. Here, different loss processes play the role of a continuous measurement. For example, it has been demonstrated that strong two-body losses give rise to an effective two-body hardcore constraint, in this way turning losses into a tool to create strong correlations~\cite{Syassen2008,GarciaRipoll2009}. For strong three-body losses, this effect has also been predicted theoretically to give rise to intriguing many-body phenomena such as dimer superfluids and -solids~\cite{Daley2009}, or the fractional quantum Hall effect~\cite{Roncaglia2010}. Another paradigmatic setup was introduced in Refs.~\cite{Barontini2013,Labouvie2016,Muellers2018}, where a local loss process is induced by shining a focused electron beam onto an atomic Bose-Einstein condensate. In particular, the QZE manifests itself in a non-monotonic behavior of the number of atoms lost from the condensate~ {\cite{Barontini2013,Labouvie2016}}: while it scales $\sim \gamma$ for a small dissipation strength $\gamma$, in the Zeno regime the inverse scaling $\sim 1/\gamma$ is obtained -- the fast scale $\gamma$ locally prevents the loss site to be entered by nearby particles (cf. Fig.~\ref{fig:fig1}). 

Although the QZE occurs here in many-body systems, the effect is understood in terms of the local, microscopic loss physics. In this work, we reveal a new incarnation of the QZE with a genuine many-body origin, induced by the interplay of strong quantum correlations, gapless modes, and a localized loss. To this end, we study a one dimensional wire of interacting fermions prepared in their ground state: this constitutes the dissipative nonequilibrium counterpart of the paradigmatic Kane-Fisher problem~\cite{Kane1992,Kane1992Long}.

We show that fluctuations strongly renormalize the loss barrier in the vicinity of the {initial} Fermi momentum $k_F$, even if it is weak on the microscopic scale. For repulsive interactions, the loss barrier is indefinitely enhanced at $k_F$ (cf. Fig.~\ref{fig:fig1}). A \emph{fluctuation-induced} QZE then manifests itself by the loss barrier becoming fully opaque for momenta $\sim k_F$. The opposite behavior with a renormalization group (RG) flow towards a transparent fixed point is observed for attractive interactions. This leads to a fluctuation-induced transparency. 

Although the defining feature of the localized dissipation is the absence of a unitarity constraint for scattering off it, unitarity is thus \emph{emerging} exactly at the Fermi level. In fact, the fixed points are analogous to the ones of Kane and Fisher~\cite{Kane1992}. However, the approach to the fixed points, i.e. the physics in the vicinity of the Fermi surface, strongly deviates from the Kane-Fisher scenario. This is highlighted for attractive interactions: here, observables scale \emph{logarithmically} with, e.g., temperature, instead of the more common algebraic behavior.

The open-system nature of the setup provides an opportunity to probe the system via its output and thus calls for suitable new observables without closed system counterpart. We show that the momentum or energy resolved escape probability turns out to be a realistic measure to detect the fluctuation-induced QZE in future experiments with ultracold atoms.

In the following, we substantiate these findings in two complementary approaches: first, within a minimalistic bosonization approach, providing a simple qualitative picture. Second, via a microscopic calculation, taking into account the dynamical nature of the open system problem and elucidating the physical mechanism behind our results.

\emph{Microscopic model  ---} We consider a wire of spinless fermions with mass $m$, interacting through a short-range potential $V(x)$, thus obeying the Hamiltonian 
\begin{equation}
\label{eq:Hamiltonian-microscopic}
 H = -\int_x \psi^\dagger(x)\frac{\nabla^2}{2m}\psi(x) + \int_{x,y} V(x-y) n(x)n(y),
\end{equation}
with $\psi^\dagger,\psi$ fermionic operators and $n(x) =\psi^\dagger(x)\psi(x)$. We assume the wire to be infinitely long, so that $\int_x = \int_{-\infty}^{+\infty}\dd x$, and to be prepared at $T=0$. At time $t=0$, a localized loss is switched on at $x=0$: this will generate particle currents, thus driving the system out of equilibrium. 
We model the loss as a localized coupling to an empty Markovian bath, thus describing the irreversible loss of atoms from the wire. Its dynamics will be then conveniently described by the quantum master equation~\cite{Zoller_book}
\begin{equation}
\label{eq:master-equation}
\partial_t \rho = - i [H,\rho ] + \int_x \Gamma(x) \left[L \rho L^\dag - \tfrac{1}{2} \{L^\dag L, \rho\}\right],
\end{equation}
with $L(x) = \psi(x)$, $\Gamma(x) = \gamma \delta(x)$, $\gamma$  being the dissipation strength.

\emph{Luttinger liquid description  ---} To obtain a first simple picture, we consider a long-wavelength description of Hamiltonian~\eqref{eq:Hamiltonian-microscopic} in terms of a Luttinger liquid~\cite{Giamarchi_book}
\begin{equation}
\label{eq:H-Luttinger}
H =  \frac{v}{2\pi} \int_x \left[ g\, (\partial_x \phi)^2 +g^{-1}\, (\partial_x \theta)^2  \right]
\end{equation}
with $v$ the speed of sound, $g$ a parameter encoding the effect of interactions, and $\theta$ and $\phi$ bosonic fields related to density and phase fluctuations, respectively.  
Given the nonequilibrium nature of the system, the usual equilibrium techniques are inadequate to treat the problem, and therefore it is convenient to resort to a Keldysh description~\cite{Kamenev_book,Sieberer_review}. 
In order to include the local loss in the bosonization language, we map the master equation~\eqref{eq:master-equation} onto a Keldysh action, and then bosonize the fields~\cite{suppmat,Mitra2011,Schiro2015, Buchhold2015}. Analogously to the case of a potential barrier~\cite{Kane1992,Kane1992Long}, this yields a local backscattering term
\begin{equation}
\label{eq:backscattering}
S_\text{back} = - 2i \gamma  \int_{x,t} \delta(x) \left( \ee^{i \phi_q} - \cos \theta_q \right)\cos \theta_c ,
\end{equation}
where the labels $c,q$ denote the classical and quantum fields, respectively~\cite{Kamenev_book}. Notice that both $\theta$ and $\phi$ appear, differently from the case of a potential barrier, where only $\theta$ is involved: the field $\phi$ accounts for the currents flowing towards the impurity. 
Following Refs.~\cite{Kane1992,Kane1992Long}, we study the renormalization of the barrier in the limiting case of weak dissipation $\gamma \to 0$. The renormalization of $\gamma$ at long wavelengths is then determined  {within a momentum-shell RG scheme~\cite{suppmat}, and} produces the flow equation
\begin{equation}
\label{eq:RG-luttinger}
\frac{\dd \gamma}{\dd \ell} = (1-g) \gamma.
\end{equation}
This entails that the particle loss is expected to be suppressed for slow
modes. For attractive interactions, the perturbation is irrelevant in the RG sense as $\gamma\to 0$ and, thus, the flow suppresses the dissipation strength. In contrast, for repulsive interactions ($g < 1$) the strength of the localized loss {is relevant in the RG sense and flows to infinity, so that losses become suppressed by the QZE.}
Eq.~\eqref{eq:RG-luttinger} is remarkably similar to the one obtained for the renormalization of a potential barrier~\cite{Kane1992,Kane1992Long}, despite the fact that the present system is subject to dissipation and is out of equilibrium. 

In order to certify the domain of validity of Eq.~\eqref{eq:RG-luttinger} during the time evolution of the system, and its effect on the observables, we will analyze directly the microscopic model in Eq.~\eqref{eq:Hamiltonian-microscopic}. Moreover, while the previous analysis is perturbative in $\gamma$, the following, complementary approach is exact in $\gamma$ and perturbative in the microscopic interaction.

\emph{Dynamical regimes ---} As the dynamics following the quench of the dissipative impurity is remarkably complex~\cite{Barmettler2011,Kepesidis2012,Vidanovic2014,Kiefer2017}, we clarify its different stages by first solving numerically the non-interacting model on a lattice with periodic boundary conditions, described by the Hamiltonian $H = -\sum_{j=1}^{L}(\psi_{j+1}^\dag \psi_j + \text{h.c.})$, with $L$ the size of the system and $\psi_j,\psi^\dagger_j$ the fermionic operators on site $j$. The characterizing parameters are the system length $L$, the initial density $n_0$, and the dissipation strength $\gamma$. 
%
\begin{figure}[t!]
\includegraphics[width=8.3cm]{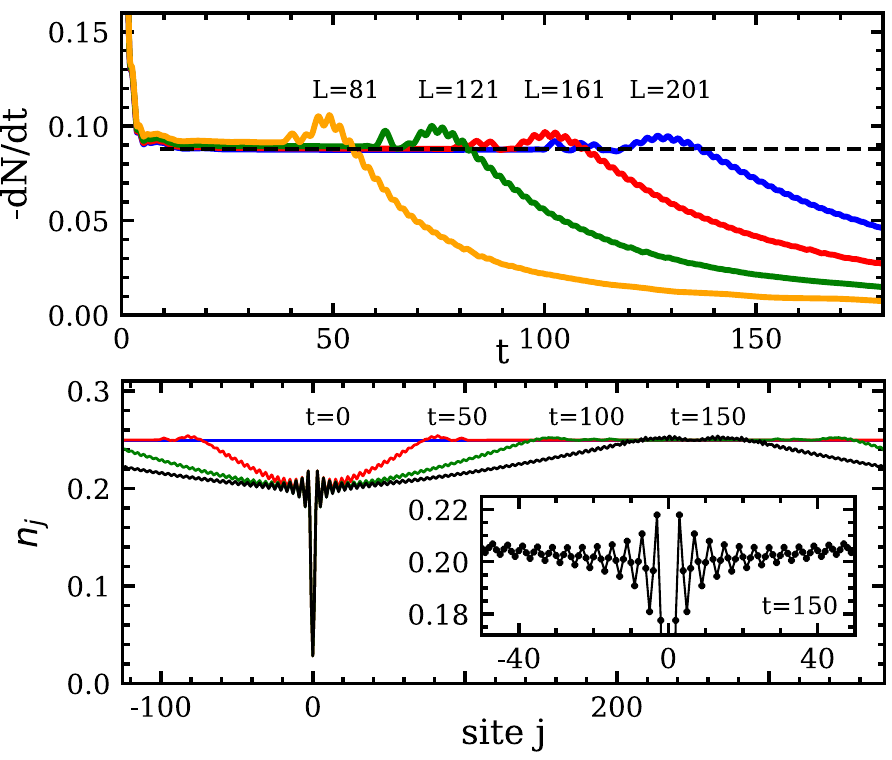}
\caption{(Color online). Upper panel: Particle loss rate from the lattice model as a function of time elapsed from the quench, for different system sizes $L$. The dashed line indicates the regime of constant loss.
Lower panel, main plot: Density profile from the lattice model {with $L=501$}, for different times elapsed from the quench. 
Inset: Friedel oscillations around the loss site. For all curves $\gamma = 3$ and $N(0)/L = 0.25 $, {with $N(t)$ the number of particles}.  }
\label{fig:fig2}
\end{figure}
%
Fig.~\ref{fig:fig2} (upper panel) shows the particle loss rate as a function of time, from which one can identify three regimes: i)~For $t < t_\text{I} \sim \gamma^{-1}$ an initial, fast depletion of particles occurs close to the loss impurity. 
ii)~For $t_\text{I} < t < t_\text{II}$, a steady particle current is established, flowing from the yet unperturbed regions of the wire at $x > vt$ (with $v$ the speed of sound) towards the loss site. 
%
%
%
This regime, which we will focus on, lasts up to a \emph{macroscopic} time $t_\text{II} \sim L$. iii)~For $t>t_\text{II}$, the entire system experiences the effect of the dissipation and the particle loss rate slows down, until the system is eventually depleted.      
In Fig.~\ref{fig:fig2} (lower panel) we show the density profile during the second regime $t_\text{I}<t<t_\text{II}$. 
At the impurity site the density is strongly depleted, while the density in the surrounding region exhibits a less pronounced depletion, which heals back to the initial value of the density on a distance $\sim v t$. 
Crucially, the density around the impurity displays Friedel oscillations, originating from the Fermi step in the momentum distribution of the initial state.     

\emph{Friedel oscillations ---} As the second dynamical regime is extensively long, we will focus on it in the remaining part of this Letter. We substantiate the existence of Friedel oscillations by an analytical solution of the non-interacting continuum problem with a localized loss. By using the Green's functions method~\cite{suppmat}, we derive the time-dependent density profile $n(x,t) = \int_k \left|G_R(x,-k,t)\right|^2\, n_0(k)$, with $G_R(x,k,t)$ the single-particle retarded Green's function, and $n_0(k) = \theta(k_F^2-k^2)$ the momentum distribution in the initial state,  {with $k_F = \pi n_0$ the Fermi momentum and $n_0$ the initial density.} 
The single-particle retarded Green's function $G_R$ is then evaluated exactly by solving the corresponding Dyson equation~\cite{Kamenev_book}, and its explicit form is reported in~\cite{suppmat}. By taking the limit $t\to \infty$,  we obtain a stationary value for $n(x)$ corresponding to the second regime discussed above: in fact, by having already taken the thermodynamic limit $L\to\infty$ we implicitly assumed $t_\text{II} \to +\infty$, so that the system is ``frozen'' in the second regime. 

We then find~\cite{suppmat} $ n(x) - n_\text{ness} \propto \sin(2k_Fx)/x $, which holds for $x \gg k_F^{-1}$, with $n_\text{ness}$ the uniform background of the stationary state.  {Remarkably, the discontinuity in the momentum distribution remains at the initial value of $k_F$~\cite{suppmat}}.
%
%
These density modulations will generate, in an interacting system, an additional barrier renormalizing the original one. In fact, for momenta close to $k_F$, the virtual scattering processes between the two barriers add up to an effective impenetrable one (for repulsive interactions) or a vanishing one (for attractive interactions)~\cite{Matveev1993,Yue1994}.
In the following, we show that this mechanism also applies to the present nonequilibrium, dissipative case.\\

%

\begin{figure}[t!]
\centering
\includegraphics[width=8.6cm]{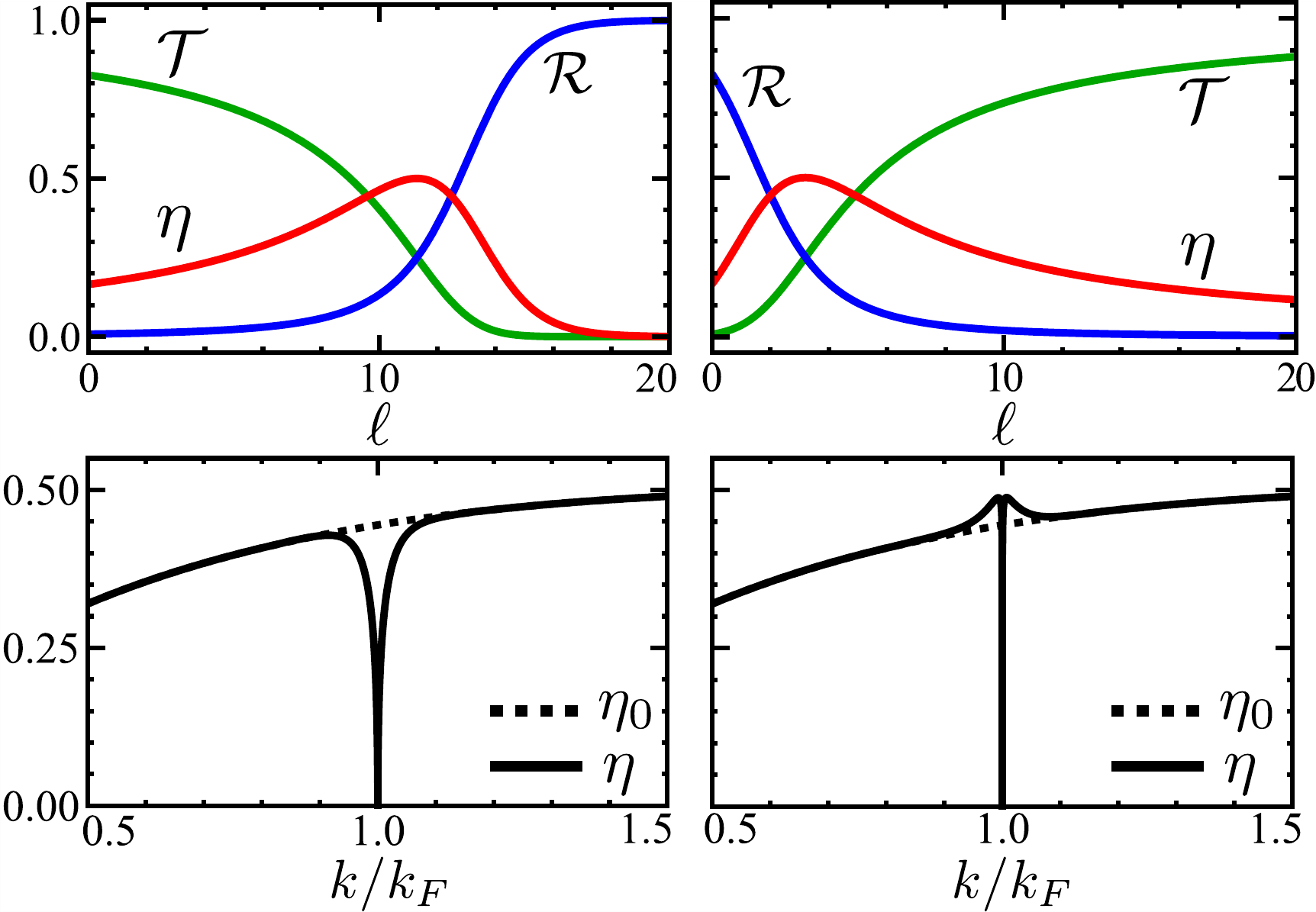}
\caption{(Color online). Upper panel: RG flow of $\mathcal{T}$, $\mathcal{R}$, and $\eta$. For $\alpha>0$  (left, $\gamma = 0.2$) a fully reflective fixed point is approached, while for $\alpha<0$ (right, $\gamma = 20$) the system is perfectly transmissive at the fixed point. 
Lower panel: Renormalized $\eta$ as a function of momentum in comparison to the non-interacting value $\eta_0$, for $\alpha > 0$ (left panel) and $\alpha < 0$ (right panel). For all curves $\gamma = 4$.}
\label{fig:fig4}
\end{figure}

\emph{Transport properties ---} To gain further insight on Eq.~\eqref{eq:RG-luttinger}, we consider the transport properties. The transmission and reflection probabilities $\mathcal{T}(k)$ and $\mathcal{R}(k)$, respectively, for a particle with momentum $k > 0$ impinging upon the dissipative impurity can be read off the retarded Green's function~\cite{suppmat}.  The loss of unitarity related to the scattering off the loss barrier is then quantified by
\begin{equation}
\eta(k) = 1 - \mathcal{T}(k) - \mathcal{R}(k).
\end{equation}
The escape probability $\eta(k)$ describes the probability that a particle with momentum $k$ is absorbed into the bath.  $\eta(k)$, which is related to the Fourier transform of $\langle \psi^\dagger(t,x=0)\psi(t',x=0)\rangle$~\cite{suppmat}, is the key quantity of the present analysis, for three reasons: i) it bears signatures of the QZE, ii) as a momentum-resolved quantity, it is sensitive to the renormalization of long-wavelength modes, and iii) it can be directly related to experimentally measurable quantities. \\
For the interactionless case $\eta_0(k) = 2\gamma v_k/(\gamma+v_k)^2$, with $v_k = |k|/m$ the group velocity, showing that losses are suppressed for both $v_k/\gamma \to 0$ and $v_k/\gamma \to \infty$.      
The perturbative corrections to $\mathcal{T}, \mathcal{R}$ due to the interaction potential $V(x)$ can be computed in analogy to equilibrium~\cite{Matveev1993,Yue1994,Aristov2010} and they yield~\cite{suppmat}
\begin{subequations}
\label{eq:perturbative-corrections}
\begin{align}
\delta\mathcal{T} & = 2\alpha\, \mathcal{T}_0 \mathcal{R}_0 \, \log|d(k-k_F)|, \\
\delta\mathcal{R} & = \alpha\, \mathcal{R}_0\left(\mathcal{R}_0 + \mathcal{T}_0 - 1\right) \, \log|d(k-k_F)|,
\end{align}
\end{subequations}
with $\mathcal{T}_0,\mathcal{R}_0$ the bare values, $d$ a length scale to be chosen as the largest between the spatial extent of the interaction $V(x)$ and the Fermi wavelength, and $\alpha = [\widetilde{V}(0)-\widetilde{V}(2k_F)]/(2\pi v_F)$, $\widetilde{V}(k)$ being the Fourier transform of $V(x)$ and $v_F\equiv|k_F|/m$ the Fermi velocity. The two contributions $\widetilde{V}(0)$ and $\widetilde{V}(2k_F)$ derive from the exchange and Hartree part of the interaction, respectively; $\alpha>0$ corresponds to repulsive interactions, and $\alpha<0$ to attractive ones.         
While the perturbative corrections~\eqref{eq:perturbative-corrections} are in principle controlled by an expansion in $\alpha\ll 1$, they actually diverge logarithmically for $k\to k_F$. 
These divergences can be resummed by an RG treatment~\cite{Matveev1993,Yue1994, suppmat}, leading to the RG flow equations
\begin{equation}
\label{eq:RG-equations}
\frac{\dd \mathcal{T}}{\dd \ell}  = - 2\alpha \, \mathcal{T} \mathcal{R} , \qquad \frac{\dd \mathcal{R}}{\dd \ell}  = - \alpha\, \mathcal{R}\left(\mathcal{R} + \mathcal{T} - 1\right),
\end{equation}
with the flow to be stopped at $\ell = -\log|d(k-k_F)|$. Eqs.~\eqref{eq:RG-equations} have one stable fixed point: $\mathcal{T}^*=0$, $\mathcal{R}^*=1$ for $\alpha>0$ and $\mathcal{T}^*=1$, $\mathcal{R}^*=0$ for $\alpha<0$. Physically, this entails that tunneling through the dissipative impurity is suppressed at $k=k_F$ for repulsive interactions, while it is maximally enhanced for attractive interactions, similarly to the case of a potential barrier~\cite{Matveev1993,Yue1994}.
However, two novel remarkable features, emerge from the solutions of Eqs.~\eqref{eq:RG-equations}. First, for both attractive and repulsive interactions, $\eta^* = 0$, implying that particles with $k=k_F$ are not emitted into the bath, but are actually ``trapped'' inside the wire, signaling emergent unitarity at the Fermi level.  
Second, $\eta(k)$ approaches its fixed point value in qualitatively different ways, depending on the sign of $\alpha$:
\begin{equation}
\label{eq:fluctuation-QZ}
\eta(k) \sim 
\begin{cases}
 |k-k_F|^{\alpha} & \text{for} \quad \alpha > 0, \\
-1/\log|d(k-k_F)| & \text{for}\quad  \alpha < 0.
\end{cases}
\end{equation}
%
%
This asymmetry, also visible in the behaviour of $\mathcal{T}(k)$ and $\mathcal{R}(k)$ for $k \to k_F $, does not occur for the case of a potential barrier, where the fixed-point values are approached algebraically in both cases~\cite{Matveev1993,Yue1994}. 
Eq.~\eqref{eq:fluctuation-QZ} is the key result of this work: the escape probability at the Fermi momentum is strongly renormalized by fluctuations, which suppress it. For $\alpha>0$, this happens as if $\gamma \to +\infty$, thus producing a QZE; for $\alpha<0$, instead, this happens as if $\gamma \to 0$, the impurity thus becoming transparent.
Fig.~\ref{fig:fig4} (upper panels) shows the RG flow of $\mathcal{T}$, $\mathcal{R}$, and $\eta$, for both repulsive and attractive interactions. The flow of $\eta$ may be non-monotonic depending on the sign of $\alpha$ and on the bare value $\eta_0$.
To make contact with the Luttinger formulation and Fig.~\ref{fig:fig1}, it is possible to reparametrize $\mathcal{T}(k)$, $\mathcal{R}(k)$ and $\eta(k)$ in terms of a single function $\widetilde{\gamma}(k)$~\cite{suppmat}. For the escape probability one finds $\eta(k)= 2\widetilde{\gamma}(k)/[1+\widetilde{\gamma}(k)]^2$. The RG flow of $\widetilde{\gamma}$ can be determined from Eqs.~\eqref{eq:RG-equations} as
\begin{equation}
\label{eq:RG-gamma}
\frac{\dd \widetilde{\gamma}}{\dd \ell} = \alpha \frac{\widetilde{\gamma}^2}{1+\widetilde{\gamma}}.
\end{equation}
The fixed points of Eq.~\eqref{eq:RG-equations} translate then to $\widetilde{\gamma}^*= \infty$ for $\alpha>0$ and $\widetilde{\gamma}^*= 0$ for $\alpha<0$. 
$\widetilde{\gamma}$ can therefore be interpreted as the effective strength of the localized dissipation, thus bridging the Luttinger result~\eqref{eq:RG-luttinger} with the one obtained from the microscopic model. In fact, for a Luttinger parameter $g \simeq 1 $, one has $ 1 - g \simeq g^{-1}-1\simeq \alpha$~\cite{Fisher_review}, and therefore Eqs.~\eqref{eq:RG-luttinger} and~\eqref{eq:RG-gamma} coincide for $\widetilde{\gamma} \gg 1$, upon the identifications $\widetilde{\gamma}\equiv\gamma$. The discrepancy for $\widetilde{\gamma} \ll 1$ can be understood as the limits $\gamma\to 0$ and $\alpha \to 0$ do not commute for  the nonequilibrium stationary state.

The behavior of $\eta(k)$ under RG can be therefore rationalized in terms of the flow of $\gamma(\ell)$ (see Fig.~\ref{fig:fig1}): $\eta$ reaches its fixed point $\eta^*=0$ either for $\gamma\to \infty$ or for $\gamma \to 0$, in the former case thus resulting in a fluctuation-induced QZE, and in the latter one a fluctuation-induced transparency. 
In Fig.~\ref{fig:fig4} (lower panels) we show the value of $\eta(k)$ as a function of $k$ reconstructed from the RG flow: its value drops to zero at $k_F$ for both attractive and repulsive interactions. As a consequence of the non-monotonicity of the RG flow, $\eta(k)$ may increase for momenta in the vicinity of $k_F$ (right panel). 

\emph{Observables ---} The fluctuation-induced QZE can be naturally detected by harnessing the energy or momentum resolved flow of particles leaving the wire, without destructive measurements. 
For illustrative purposes, we consider the following model, inspired by the input-output formalism of quantum optics~\cite{Zoller_book}. 
The local dissipation originates from the wire being coupled to a continuum of fermionic modes outside the wire. For definiteness, we assume particles to exit the wire by expanding isotropically in the surrounding vacuum, which could, e.g., be realized in the setup of  Ref.~\cite{Lebrat2018} by a local transfer to an untrapped internal state. The particles could be described by operators $c_\qq, c^\dagger_\qq$, with $\qq$ a three-dimensional momentum. These modes are coupled to the wire at $x = 0$ through the Hamiltonian $H_\text{int} = \sum_{\qq}(g_\qq c^\dagger_\qq \psi_0+ \text{h.c.})$, with $g_\qq$ the coupling of the $\qq$-mode to the wire and $\psi_0 \equiv \psi(x=0)$.
In the second regime, the constant rate of particles with momentum $\qq$  {and energy $\epsilon_\qq$} leaving the wire is then given by~\cite{suppmat}
\begin{equation}
\label{eq:particle-production}
\frac{\dd \langle c^\dagger_\qq c_\qq\rangle}{\dd t} =  \theta(E_F-\epsilon_\qq) \frac{|g_\qq|^2}{\gamma}  \eta(\epsilon_\qq),
\end{equation}
with $E_F = k_F^2/2m$, thus providing a connection between $\eta$ and an experimentally accessible quantity. The bare Fermi distribution enters Eq.~\eqref{eq:particle-production} as the approach is perturbative in the interaction: it is expected to be smeared out by stronger interactions~\cite{Giamarchi_book}. \\

\emph{Finite temperature and size ---} The unavoidable presence of a finite temperature $T$ and system size $L$ in realistic systems can be accounted for by our RG analysis, and their variation actually leveraged to disclose the novel collective behaviors described above. In fact, a finite $T$ (resp. $L$) cuts off the RG flow at a scale $\ell_T =-\log(d\,  T)$ (resp. $\ell_L = \log(L/d)$). As a consequence $\eta(k_F)$ can exhibit a non-monotonic behaviour as a function of the considered length-scale (cf. Fig.~\ref{fig:fig4}, upper panels). For instance, the value of $\eta(k_F)$ \emph{increases} up to $\sim$ 100\% by \emph{reducing} the temperature of a gas from $T_\text{Fermi}$ to $0.1T_\text{Fermi}$, thus suggesting that the effects above discussed are observable within the current experimental setups~\cite{Brantut_review,Lebrat2018}.
Although in nonequilibrium systems the (effective) temperature may change during the RG flow~\cite{Mitra2011,DellaTorre2012,Schiro2014}, we argue that this is not the case in our setup. In fact, we expect the temperature to be enforced by the extensive ``reservoir'' constituted by the far ends of the wire, rather than by the impurity, which is a local perturbation. A quantitative answer requires to extend our RG analysis to two loops, which we leave to future work.

\emph{Conclusions ---} We have shown that a one dimensional ultracold gas of fermions displays novel many-body effects in presence of a localized loss. Loss is suppressed close to the Fermi energy, effectively restoring unitarity. This consequence of the renormalization of the dissipation strength can be interpreted as an incarnation of the QZE. Moreover, transport properties are modified similarly to the case of a potential barrier. 
These effects would be experimentally accessible by analyzing the ejected flow of particles energy or momentum resolved, without further destructive measurements. 
An analogous situation to the considered stationary regime could be obtained in experiments with systems coupled to reservoirs at both ends~\cite{suppmat} (cf. Ref.~\cite{Lebrat2018}) .
%
%

\emph{Acknowledgements ---} We thank J.-P.~Brantut, L.~Corman, J.~Marino, A.~Rosch and F.~Tonielli for useful discussions. We acknowledge support by the Institutional Strategy of the University of Cologne within the German Excellence Initiative (ZUK 81), by the funding from the European Research Council (ERC) under the Horizon 2020 research and innovation program, Grant Agreement No. 647434 (DOQS) and No. 648166 (Phonton), and by the DFG  {(TR 185 projects B3 and B4, Collaborative Research Center (CRC) 1238 Project No. 277146847 -  projects C04 and C05)}. This research was supported in part by the National Science Foundation under Grant No. NSF PHY-1748958.

\bibliography{biblio}

\begin{thebibliography}{30}%
\makeatletter
\providecommand \@ifxundefined [1]{%
 \@ifx{#1\undefined}
}%
\providecommand \@ifnum [1]{%
 \ifnum #1\expandafter \@firstoftwo
 \else \expandafter \@secondoftwo
 \fi
}%
\providecommand \@ifx [1]{%
 \ifx #1\expandafter \@firstoftwo
 \else \expandafter \@secondoftwo
 \fi
}%
\providecommand \natexlab [1]{#1}%
\providecommand \enquote  [1]{``#1''}%
\providecommand \bibnamefont  [1]{#1}%
\providecommand \bibfnamefont [1]{#1}%
\providecommand \citenamefont [1]{#1}%
\providecommand \href@noop [0]{\@secondoftwo}%
\providecommand \href [0]{\begingroup \@sanitize@url \@href}%
\providecommand \@href[1]{\@@startlink{#1}\@@href}%
\providecommand \@@href[1]{\endgroup#1\@@endlink}%
\providecommand \@sanitize@url [0]{\catcode `\\12\catcode `\$12\catcode
  `\&12\catcode `\#12\catcode `\^12\catcode `\_12\catcode `\%12\relax}%
\providecommand \@@startlink[1]{}%
\providecommand \@@endlink[0]{}%
\providecommand \url  [0]{\begingroup\@sanitize@url \@url }%
\providecommand \@url [1]{\endgroup\@href {#1}{\urlprefix }}%
\providecommand \urlprefix  [0]{URL }%
\providecommand \Eprint [0]{\href }%
\providecommand \doibase [0]{http://dx.doi.org/}%
\providecommand \selectlanguage [0]{\@gobble}%
\providecommand \bibinfo  [0]{\@secondoftwo}%
\providecommand \bibfield  [0]{\@secondoftwo}%
\providecommand \translation [1]{[#1]}%
\providecommand \BibitemOpen [0]{}%
\providecommand \bibitemStop [0]{}%
\providecommand \bibitemNoStop [0]{.\EOS\space}%
\providecommand \EOS [0]{\spacefactor3000\relax}%
\providecommand \BibitemShut  [1]{\csname bibitem#1\endcsname}%
\let\auto@bib@innerbib\@empty
\bibitem [{\citenamefont {Misra}\ and\ \citenamefont
  {Sudarshan}(1977)}]{Misra1977}%
  \BibitemOpen
  \bibfield  {author} {\bibinfo {author} {\bibfnamefont {B.}~\bibnamefont
  {Misra}}\ and\ \bibinfo {author} {\bibfnamefont {E.~C.~G.}\ \bibnamefont
  {Sudarshan}},\ }\href {\doibase 10.1063/1.523304} {\bibfield  {journal}
  {\bibinfo  {journal} {J. Math. Phys.}\ }\textbf {\bibinfo {volume} {18}},\
  \bibinfo {pages} {756} (\bibinfo {year} {1977})}\BibitemShut {NoStop}%
\bibitem [{\citenamefont {Syassen}\ \emph {et~al.}(2008)\citenamefont
  {Syassen}, \citenamefont {Bauer}, \citenamefont {Lettner}, \citenamefont
  {Volz}, \citenamefont {Dietze}, \citenamefont {Garc{\'\i}a-Ripoll},
  \citenamefont {Cirac}, \citenamefont {Rempe},\ and\ \citenamefont
  {D{\"u}rr}}]{Syassen2008}%
  \BibitemOpen
  \bibfield  {author} {\bibinfo {author} {\bibfnamefont {N.}~\bibnamefont
  {Syassen}}, \bibinfo {author} {\bibfnamefont {D.~M.}\ \bibnamefont {Bauer}},
  \bibinfo {author} {\bibfnamefont {M.}~\bibnamefont {Lettner}}, \bibinfo
  {author} {\bibfnamefont {T.}~\bibnamefont {Volz}}, \bibinfo {author}
  {\bibfnamefont {D.}~\bibnamefont {Dietze}}, \bibinfo {author} {\bibfnamefont
  {J.~J.}\ \bibnamefont {Garc{\'\i}a-Ripoll}}, \bibinfo {author} {\bibfnamefont
  {J.~I.}\ \bibnamefont {Cirac}}, \bibinfo {author} {\bibfnamefont
  {G.}~\bibnamefont {Rempe}}, \ and\ \bibinfo {author} {\bibfnamefont
  {S.}~\bibnamefont {D{\"u}rr}},\ }\href {\doibase 10.1126/science.1155309}
  {\bibfield  {journal} {\bibinfo  {journal} {Science}\ }\textbf {\bibinfo
  {volume} {320}},\ \bibinfo {pages} {1329} (\bibinfo {year}
  {2008})}\BibitemShut {NoStop}%
\bibitem [{\citenamefont {Garc{\'\i}­a-Ripoll}\ \emph
  {et~al.}(2009)\citenamefont {Garc{\'\i}­a-Ripoll}, \citenamefont {D\"{u}rr},
  \citenamefont {Syassen}, \citenamefont {Bauer}, \citenamefont {Lettner},
  \citenamefont {Rempe},\ and\ \citenamefont {Cirac}}]{GarciaRipoll2009}%
  \BibitemOpen
  \bibfield  {author} {\bibinfo {author} {\bibfnamefont {J.~J.}\ \bibnamefont
  {Garc{\'\i}­a-Ripoll}}, \bibinfo {author} {\bibfnamefont {S.}~\bibnamefont
  {D\"{u}rr}}, \bibinfo {author} {\bibfnamefont {N.}~\bibnamefont {Syassen}},
  \bibinfo {author} {\bibfnamefont {D.~M.}\ \bibnamefont {Bauer}}, \bibinfo
  {author} {\bibfnamefont {M.}~\bibnamefont {Lettner}}, \bibinfo {author}
  {\bibfnamefont {G.}~\bibnamefont {Rempe}}, \ and\ \bibinfo {author}
  {\bibfnamefont {J.~I.}\ \bibnamefont {Cirac}},\ }\href
  {http://stacks.iop.org/1367-2630/11/i=1/a=013053} {\bibfield  {journal}
  {\bibinfo  {journal} {New J. Phys.}\ }\textbf {\bibinfo {volume} {11}},\
  \bibinfo {pages} {013053} (\bibinfo {year} {2009})}\BibitemShut {NoStop}%
\bibitem [{\citenamefont {Daley}\ \emph {et~al.}(2009)\citenamefont {Daley},
  \citenamefont {Taylor}, \citenamefont {Diehl}, \citenamefont {Baranov},\ and\
  \citenamefont {Zoller}}]{Daley2009}%
  \BibitemOpen
  \bibfield  {author} {\bibinfo {author} {\bibfnamefont {A.~J.}\ \bibnamefont
  {Daley}}, \bibinfo {author} {\bibfnamefont {J.~M.}\ \bibnamefont {Taylor}},
  \bibinfo {author} {\bibfnamefont {S.}~\bibnamefont {Diehl}}, \bibinfo
  {author} {\bibfnamefont {M.}~\bibnamefont {Baranov}}, \ and\ \bibinfo
  {author} {\bibfnamefont {P.}~\bibnamefont {Zoller}},\ }\href {\doibase
  10.1103/PhysRevLett.102.040402} {\bibfield  {journal} {\bibinfo  {journal}
  {Phys. Rev. Lett.}\ }\textbf {\bibinfo {volume} {102}},\ \bibinfo {pages}
  {040402} (\bibinfo {year} {2009})}\BibitemShut {NoStop}%
\bibitem [{\citenamefont {Roncaglia}\ \emph {et~al.}(2010)\citenamefont
  {Roncaglia}, \citenamefont {Rizzi},\ and\ \citenamefont
  {Cirac}}]{Roncaglia2010}%
  \BibitemOpen
  \bibfield  {author} {\bibinfo {author} {\bibfnamefont {M.}~\bibnamefont
  {Roncaglia}}, \bibinfo {author} {\bibfnamefont {M.}~\bibnamefont {Rizzi}}, \
  and\ \bibinfo {author} {\bibfnamefont {J.~I.}\ \bibnamefont {Cirac}},\ }\href
  {\doibase 10.1103/PhysRevLett.104.096803} {\bibfield  {journal} {\bibinfo
  {journal} {Phys. Rev. Lett.}\ }\textbf {\bibinfo {volume} {104}},\ \bibinfo
  {pages} {096803} (\bibinfo {year} {2010})}\BibitemShut {NoStop}%
\bibitem [{\citenamefont {Barontini}\ \emph {et~al.}(2013)\citenamefont
  {Barontini}, \citenamefont {Labouvie}, \citenamefont {Stubenrauch},
  \citenamefont {Vogler}, \citenamefont {Guarrera},\ and\ \citenamefont
  {Ott}}]{Barontini2013}%
  \BibitemOpen
  \bibfield  {author} {\bibinfo {author} {\bibfnamefont {G.}~\bibnamefont
  {Barontini}}, \bibinfo {author} {\bibfnamefont {R.}~\bibnamefont {Labouvie}},
  \bibinfo {author} {\bibfnamefont {F.}~\bibnamefont {Stubenrauch}}, \bibinfo
  {author} {\bibfnamefont {A.}~\bibnamefont {Vogler}}, \bibinfo {author}
  {\bibfnamefont {V.}~\bibnamefont {Guarrera}}, \ and\ \bibinfo {author}
  {\bibfnamefont {H.}~\bibnamefont {Ott}},\ }\href {\doibase
  10.1103/PhysRevLett.110.035302} {\bibfield  {journal} {\bibinfo  {journal}
  {Phys. Rev. Lett.}\ }\textbf {\bibinfo {volume} {110}},\ \bibinfo {pages}
  {035302} (\bibinfo {year} {2013})}\BibitemShut {NoStop}%
\bibitem [{\citenamefont {Labouvie}\ \emph {et~al.}(2016)\citenamefont
  {Labouvie}, \citenamefont {Santra}, \citenamefont {Heun},\ and\ \citenamefont
  {Ott}}]{Labouvie2016}%
  \BibitemOpen
  \bibfield  {author} {\bibinfo {author} {\bibfnamefont {R.}~\bibnamefont
  {Labouvie}}, \bibinfo {author} {\bibfnamefont {B.}~\bibnamefont {Santra}},
  \bibinfo {author} {\bibfnamefont {S.}~\bibnamefont {Heun}}, \ and\ \bibinfo
  {author} {\bibfnamefont {H.}~\bibnamefont {Ott}},\ }\href {\doibase
  10.1103/PhysRevLett.116.235302} {\bibfield  {journal} {\bibinfo  {journal}
  {Phys. Rev. Lett.}\ }\textbf {\bibinfo {volume} {116}},\ \bibinfo {pages}
  {235302} (\bibinfo {year} {2016})}\BibitemShut {NoStop}%
\bibitem [{\citenamefont {M{\"u}llers}\ \emph {et~al.}(2018)\citenamefont
  {M{\"u}llers}, \citenamefont {Santra}, \citenamefont {Baals}, \citenamefont
  {Jiang}, \citenamefont {Benary}, \citenamefont {Labouvie}, \citenamefont
  {Zezyulin}, \citenamefont {Konotop},\ and\ \citenamefont
  {Ott}}]{Muellers2018}%
  \BibitemOpen
  \bibfield  {author} {\bibinfo {author} {\bibfnamefont {A.}~\bibnamefont
  {M{\"u}llers}}, \bibinfo {author} {\bibfnamefont {B.}~\bibnamefont {Santra}},
  \bibinfo {author} {\bibfnamefont {C.}~\bibnamefont {Baals}}, \bibinfo
  {author} {\bibfnamefont {J.}~\bibnamefont {Jiang}}, \bibinfo {author}
  {\bibfnamefont {J.}~\bibnamefont {Benary}}, \bibinfo {author} {\bibfnamefont
  {R.}~\bibnamefont {Labouvie}}, \bibinfo {author} {\bibfnamefont {D.~A.}\
  \bibnamefont {Zezyulin}}, \bibinfo {author} {\bibfnamefont {V.~V.}\
  \bibnamefont {Konotop}}, \ and\ \bibinfo {author} {\bibfnamefont
  {H.}~\bibnamefont {Ott}},\ }\href@noop {} {\bibfield  {journal} {\bibinfo
  {journal} {Sci. Adv.}\ }\textbf {\bibinfo {volume} {4}},\ \bibinfo {pages}
  {eaat6539} (\bibinfo {year} {2018})}\BibitemShut {NoStop}%
\bibitem [{\citenamefont {Kane}\ and\ \citenamefont
  {Fisher}(1992{\natexlab{a}})}]{Kane1992}%
  \BibitemOpen
  \bibfield  {author} {\bibinfo {author} {\bibfnamefont {C.~L.}\ \bibnamefont
  {Kane}}\ and\ \bibinfo {author} {\bibfnamefont {M.~P.~A.}\ \bibnamefont
  {Fisher}},\ }\href {\doibase 10.1103/PhysRevLett.68.1220} {\bibfield
  {journal} {\bibinfo  {journal} {Phys. Rev. Lett.}\ }\textbf {\bibinfo
  {volume} {68}},\ \bibinfo {pages} {1220} (\bibinfo {year}
  {1992}{\natexlab{a}})}\BibitemShut {NoStop}%
\bibitem [{\citenamefont {Kane}\ and\ \citenamefont
  {Fisher}(1992{\natexlab{b}})}]{Kane1992Long}%
  \BibitemOpen
  \bibfield  {author} {\bibinfo {author} {\bibfnamefont {C.~L.}\ \bibnamefont
  {Kane}}\ and\ \bibinfo {author} {\bibfnamefont {M.~P.~A.}\ \bibnamefont
  {Fisher}},\ }\href {\doibase 10.1103/PhysRevB.46.15233} {\bibfield  {journal}
  {\bibinfo  {journal} {Phys. Rev. B}\ }\textbf {\bibinfo {volume} {46}},\
  \bibinfo {pages} {15233} (\bibinfo {year} {1992}{\natexlab{b}})}\BibitemShut
  {NoStop}%
\bibitem [{\citenamefont {Gardiner}\ and\ \citenamefont
  {Zoller}(2000)}]{Zoller_book}%
  \BibitemOpen
  \bibfield  {author} {\bibinfo {author} {\bibfnamefont {C.}~\bibnamefont
  {Gardiner}}\ and\ \bibinfo {author} {\bibfnamefont {P.}~\bibnamefont
  {Zoller}},\ }\href {https://books.google.de/books?id=4bJ6MgEACAAJ} {\emph
  {\bibinfo {title} {Quantum Noise}}}\ (\bibinfo  {publisher} {Springer},\
  \bibinfo {year} {2000})\BibitemShut {NoStop}%
\bibitem [{\citenamefont {Giamarchi}(2004)}]{Giamarchi_book}%
  \BibitemOpen
  \bibfield  {author} {\bibinfo {author} {\bibfnamefont {T.}~\bibnamefont
  {Giamarchi}},\ }\href {https://books.google.de/books?id=1MwTDAAAQBAJ} {\emph
  {\bibinfo {title} {Quantum Physics in One Dimension}}}\ (\bibinfo
  {publisher} {Oxford University Press},\ \bibinfo {year} {2004})\BibitemShut
  {NoStop}%
\bibitem [{\citenamefont {Kamenev}(2011)}]{Kamenev_book}%
  \BibitemOpen
  \bibfield  {author} {\bibinfo {author} {\bibfnamefont {A.}~\bibnamefont
  {Kamenev}},\ }\href {\doibase 10.1017/CBO9781139003667} {\emph {\bibinfo
  {title} {Field Theory of Non-Equilibrium Systems}}}\ (\bibinfo  {publisher}
  {Cambridge University Press},\ \bibinfo {year} {2011})\BibitemShut {NoStop}%
\bibitem [{\citenamefont {Sieberer}\ \emph {et~al.}(2016)\citenamefont
  {Sieberer}, \citenamefont {Buchhold},\ and\ \citenamefont
  {Diehl}}]{Sieberer_review}%
  \BibitemOpen
  \bibfield  {author} {\bibinfo {author} {\bibfnamefont {L.~M.}\ \bibnamefont
  {Sieberer}}, \bibinfo {author} {\bibfnamefont {M.}~\bibnamefont {Buchhold}},
  \ and\ \bibinfo {author} {\bibfnamefont {S.}~\bibnamefont {Diehl}},\ }\href
  {http://stacks.iop.org/0034-4885/79/i=9/a=096001} {\bibfield  {journal}
  {\bibinfo  {journal} {Rep. Prog. Phys.}\ }\textbf {\bibinfo {volume} {79}},\
  \bibinfo {pages} {096001} (\bibinfo {year} {2016})}\BibitemShut {NoStop}%
\bibitem [{sup()}]{suppmat}%
  \BibitemOpen
  \href@noop {} {}\bibinfo {note} {See Supplemental Material.}\BibitemShut
  {Stop}%
\bibitem [{\citenamefont {Mitra}\ and\ \citenamefont
  {Giamarchi}(2011)}]{Mitra2011}%
  \BibitemOpen
  \bibfield  {author} {\bibinfo {author} {\bibfnamefont {A.}~\bibnamefont
  {Mitra}}\ and\ \bibinfo {author} {\bibfnamefont {T.}~\bibnamefont
  {Giamarchi}},\ }\href {\doibase 10.1103/PhysRevLett.107.150602} {\bibfield
  {journal} {\bibinfo  {journal} {Phys. Rev. Lett.}\ }\textbf {\bibinfo
  {volume} {107}},\ \bibinfo {pages} {150602} (\bibinfo {year}
  {2011})}\BibitemShut {NoStop}%
\bibitem [{\citenamefont {Schir\'o}\ and\ \citenamefont
  {Mitra}(2015)}]{Schiro2015}%
  \BibitemOpen
  \bibfield  {author} {\bibinfo {author} {\bibfnamefont {M.}~\bibnamefont
  {Schir\'o}}\ and\ \bibinfo {author} {\bibfnamefont {A.}~\bibnamefont
  {Mitra}},\ }\href {\doibase 10.1103/PhysRevB.91.235126} {\bibfield  {journal}
  {\bibinfo  {journal} {Phys. Rev. B}\ }\textbf {\bibinfo {volume} {91}},\
  \bibinfo {pages} {235126} (\bibinfo {year} {2015})}\BibitemShut {NoStop}%
\bibitem [{\citenamefont {Buchhold}\ and\ \citenamefont
  {Diehl}(2015)}]{Buchhold2015}%
  \BibitemOpen
  \bibfield  {author} {\bibinfo {author} {\bibfnamefont {M.}~\bibnamefont
  {Buchhold}}\ and\ \bibinfo {author} {\bibfnamefont {S.}~\bibnamefont
  {Diehl}},\ }\href {\doibase 10.1103/PhysRevA.92.013603} {\bibfield  {journal}
  {\bibinfo  {journal} {Phys. Rev. A}\ }\textbf {\bibinfo {volume} {92}},\
  \bibinfo {pages} {013603} (\bibinfo {year} {2015})}\BibitemShut {NoStop}%
\bibitem [{\citenamefont {Barmettler}\ and\ \citenamefont
  {Kollath}(2011)}]{Barmettler2011}%
  \BibitemOpen
  \bibfield  {author} {\bibinfo {author} {\bibfnamefont {P.}~\bibnamefont
  {Barmettler}}\ and\ \bibinfo {author} {\bibfnamefont {C.}~\bibnamefont
  {Kollath}},\ }\href {\doibase 10.1103/PhysRevA.84.041606} {\bibfield
  {journal} {\bibinfo  {journal} {Phys. Rev. A}\ }\textbf {\bibinfo {volume}
  {84}},\ \bibinfo {pages} {041606} (\bibinfo {year} {2011})}\BibitemShut
  {NoStop}%
\bibitem [{\citenamefont {Kepesidis}\ and\ \citenamefont
  {Hartmann}(2012)}]{Kepesidis2012}%
  \BibitemOpen
  \bibfield  {author} {\bibinfo {author} {\bibfnamefont {K.~V.}\ \bibnamefont
  {Kepesidis}}\ and\ \bibinfo {author} {\bibfnamefont {M.~J.}\ \bibnamefont
  {Hartmann}},\ }\href {\doibase 10.1103/PhysRevA.85.063620} {\bibfield
  {journal} {\bibinfo  {journal} {Phys. Rev. A}\ }\textbf {\bibinfo {volume}
  {85}},\ \bibinfo {pages} {063620} (\bibinfo {year} {2012})}\BibitemShut
  {NoStop}%
\bibitem [{\citenamefont {Vidanovi\ifmmode~\acute{c}\else \'{c}\fi{}}\ \emph
  {et~al.}(2014)\citenamefont {Vidanovi\ifmmode~\acute{c}\else \'{c}\fi{}},
  \citenamefont {Cocks},\ and\ \citenamefont {Hofstetter}}]{Vidanovic2014}%
  \BibitemOpen
  \bibfield  {author} {\bibinfo {author} {\bibfnamefont {I.}~\bibnamefont
  {Vidanovi\ifmmode~\acute{c}\else \'{c}\fi{}}}, \bibinfo {author}
  {\bibfnamefont {D.}~\bibnamefont {Cocks}}, \ and\ \bibinfo {author}
  {\bibfnamefont {W.}~\bibnamefont {Hofstetter}},\ }\href {\doibase
  10.1103/PhysRevA.89.053614} {\bibfield  {journal} {\bibinfo  {journal} {Phys.
  Rev. A}\ }\textbf {\bibinfo {volume} {89}},\ \bibinfo {pages} {053614}
  (\bibinfo {year} {2014})}\BibitemShut {NoStop}%
\bibitem [{\citenamefont {Kiefer-Emmanouilidis}\ and\ \citenamefont
  {Sirker}(2017)}]{Kiefer2017}%
  \BibitemOpen
  \bibfield  {author} {\bibinfo {author} {\bibfnamefont {M.}~\bibnamefont
  {Kiefer-Emmanouilidis}}\ and\ \bibinfo {author} {\bibfnamefont
  {J.}~\bibnamefont {Sirker}},\ }\href {\doibase 10.1103/PhysRevA.96.063625}
  {\bibfield  {journal} {\bibinfo  {journal} {Phys. Rev. A}\ }\textbf {\bibinfo
  {volume} {96}},\ \bibinfo {pages} {063625} (\bibinfo {year}
  {2017})}\BibitemShut {NoStop}%
\bibitem [{\citenamefont {Matveev}\ \emph {et~al.}(1993)\citenamefont
  {Matveev}, \citenamefont {Yue},\ and\ \citenamefont {Glazman}}]{Matveev1993}%
  \BibitemOpen
  \bibfield  {author} {\bibinfo {author} {\bibfnamefont {K.~A.}\ \bibnamefont
  {Matveev}}, \bibinfo {author} {\bibfnamefont {D.}~\bibnamefont {Yue}}, \ and\
  \bibinfo {author} {\bibfnamefont {L.~I.}\ \bibnamefont {Glazman}},\ }\href
  {\doibase 10.1103/PhysRevLett.71.3351} {\bibfield  {journal} {\bibinfo
  {journal} {Phys. Rev. Lett.}\ }\textbf {\bibinfo {volume} {71}},\ \bibinfo
  {pages} {3351} (\bibinfo {year} {1993})}\BibitemShut {NoStop}%
\bibitem [{\citenamefont {Yue}\ \emph {et~al.}(1994)\citenamefont {Yue},
  \citenamefont {Glazman},\ and\ \citenamefont {Matveev}}]{Yue1994}%
  \BibitemOpen
  \bibfield  {author} {\bibinfo {author} {\bibfnamefont {D.}~\bibnamefont
  {Yue}}, \bibinfo {author} {\bibfnamefont {L.~I.}\ \bibnamefont {Glazman}}, \
  and\ \bibinfo {author} {\bibfnamefont {K.~A.}\ \bibnamefont {Matveev}},\
  }\href {\doibase 10.1103/PhysRevB.49.1966} {\bibfield  {journal} {\bibinfo
  {journal} {Phys. Rev. B}\ }\textbf {\bibinfo {volume} {49}},\ \bibinfo
  {pages} {1966} (\bibinfo {year} {1994})}\BibitemShut {NoStop}%
\bibitem [{\citenamefont {Aristov}\ \emph {et~al.}(2010)\citenamefont
  {Aristov}, \citenamefont {Dmitriev}, \citenamefont {Gornyi}, \citenamefont
  {Kachorovskii}, \citenamefont {Polyakov},\ and\ \citenamefont
  {W\"olfle}}]{Aristov2010}%
  \BibitemOpen
  \bibfield  {author} {\bibinfo {author} {\bibfnamefont {D.~N.}\ \bibnamefont
  {Aristov}}, \bibinfo {author} {\bibfnamefont {A.~P.}\ \bibnamefont
  {Dmitriev}}, \bibinfo {author} {\bibfnamefont {I.~V.}\ \bibnamefont
  {Gornyi}}, \bibinfo {author} {\bibfnamefont {V.~Y.}\ \bibnamefont
  {Kachorovskii}}, \bibinfo {author} {\bibfnamefont {D.~G.}\ \bibnamefont
  {Polyakov}}, \ and\ \bibinfo {author} {\bibfnamefont {P.}~\bibnamefont
  {W\"olfle}},\ }\href {\doibase 10.1103/PhysRevLett.105.266404} {\bibfield
  {journal} {\bibinfo  {journal} {Phys. Rev. Lett.}\ }\textbf {\bibinfo
  {volume} {105}},\ \bibinfo {pages} {266404} (\bibinfo {year}
  {2010})}\BibitemShut {NoStop}%
\bibitem [{\citenamefont {Fisher}\ and\ \citenamefont
  {Glazman}(1997)}]{Fisher_review}%
  \BibitemOpen
  \bibfield  {author} {\bibinfo {author} {\bibfnamefont {M.~P.~A.}\
  \bibnamefont {Fisher}}\ and\ \bibinfo {author} {\bibfnamefont {L.~I.}\
  \bibnamefont {Glazman}},\ }in\ \href {\doibase 10.1007/978-94-015-8839-3_9}
  {\emph {\bibinfo {booktitle} {Mesoscopic Electron Transport}}},\ \bibinfo
  {editor} {edited by\ \bibinfo {editor} {\bibfnamefont {L.~L.}\ \bibnamefont
  {Sohn}}, \bibinfo {editor} {\bibfnamefont {L.~P.}\ \bibnamefont
  {Kouwenhoven}}, \ and\ \bibinfo {editor} {\bibfnamefont {G.}~\bibnamefont
  {Sch{\"o}n}}}\ (\bibinfo  {publisher} {Springer Netherlands},\ \bibinfo
  {year} {1997})\BibitemShut {NoStop}%
\bibitem [{\citenamefont {Lebrat}\ \emph {et~al.}(2018)\citenamefont {Lebrat},
  \citenamefont {Gri\ifmmode~\check{s}\else \v{s}\fi{}ins}, \citenamefont
  {Husmann}, \citenamefont {H\"ausler}, \citenamefont {Corman}, \citenamefont
  {Giamarchi}, \citenamefont {Brantut},\ and\ \citenamefont
  {Esslinger}}]{Lebrat2018}%
  \BibitemOpen
  \bibfield  {author} {\bibinfo {author} {\bibfnamefont {M.}~\bibnamefont
  {Lebrat}}, \bibinfo {author} {\bibfnamefont {P.}~\bibnamefont
  {Gri\ifmmode~\check{s}\else \v{s}\fi{}ins}}, \bibinfo {author} {\bibfnamefont
  {D.}~\bibnamefont {Husmann}}, \bibinfo {author} {\bibfnamefont
  {S.}~\bibnamefont {H\"ausler}}, \bibinfo {author} {\bibfnamefont
  {L.}~\bibnamefont {Corman}}, \bibinfo {author} {\bibfnamefont
  {T.}~\bibnamefont {Giamarchi}}, \bibinfo {author} {\bibfnamefont {J.-P.}\
  \bibnamefont {Brantut}}, \ and\ \bibinfo {author} {\bibfnamefont
  {T.}~\bibnamefont {Esslinger}},\ }\href {\doibase 10.1103/PhysRevX.8.011053}
  {\bibfield  {journal} {\bibinfo  {journal} {Phys. Rev. X}\ }\textbf {\bibinfo
  {volume} {8}},\ \bibinfo {pages} {011053} (\bibinfo {year}
  {2018})}\BibitemShut {NoStop}%
\bibitem [{\citenamefont {Krinner}\ \emph {et~al.}(2017)\citenamefont
  {Krinner}, \citenamefont {Esslinger},\ and\ \citenamefont
  {Brantut}}]{Brantut_review}%
  \BibitemOpen
  \bibfield  {author} {\bibinfo {author} {\bibfnamefont {S.}~\bibnamefont
  {Krinner}}, \bibinfo {author} {\bibfnamefont {T.}~\bibnamefont {Esslinger}},
  \ and\ \bibinfo {author} {\bibfnamefont {J.-P.}\ \bibnamefont {Brantut}},\
  }\href {http://stacks.iop.org/0953-8984/29/i=34/a=343003} {\bibfield
  {journal} {\bibinfo  {journal} {J. Phys.: Condens. Matter}\ }\textbf
  {\bibinfo {volume} {29}},\ \bibinfo {pages} {343003} (\bibinfo {year}
  {2017})}\BibitemShut {NoStop}%
\bibitem [{\citenamefont {Dalla~Torre}\ \emph {et~al.}(2012)\citenamefont
  {Dalla~Torre}, \citenamefont {Demler}, \citenamefont {Giamarchi},\ and\
  \citenamefont {Altman}}]{DellaTorre2012}%
  \BibitemOpen
  \bibfield  {author} {\bibinfo {author} {\bibfnamefont {E.~G.}\ \bibnamefont
  {Dalla~Torre}}, \bibinfo {author} {\bibfnamefont {E.}~\bibnamefont {Demler}},
  \bibinfo {author} {\bibfnamefont {T.}~\bibnamefont {Giamarchi}}, \ and\
  \bibinfo {author} {\bibfnamefont {E.}~\bibnamefont {Altman}},\ }\href
  {\doibase 10.1103/PhysRevB.85.184302} {\bibfield  {journal} {\bibinfo
  {journal} {Phys. Rev. B}\ }\textbf {\bibinfo {volume} {85}},\ \bibinfo
  {pages} {184302} (\bibinfo {year} {2012})}\BibitemShut {NoStop}%
\bibitem [{\citenamefont {Schir\'o}\ and\ \citenamefont
  {Mitra}(2014)}]{Schiro2014}%
  \BibitemOpen
  \bibfield  {author} {\bibinfo {author} {\bibfnamefont {M.}~\bibnamefont
  {Schir\'o}}\ and\ \bibinfo {author} {\bibfnamefont {A.}~\bibnamefont
  {Mitra}},\ }\href {\doibase 10.1103/PhysRevLett.112.246401} {\bibfield
  {journal} {\bibinfo  {journal} {Phys. Rev. Lett.}\ }\textbf {\bibinfo
  {volume} {112}},\ \bibinfo {pages} {246401} (\bibinfo {year}
  {2014})}\BibitemShut {NoStop}%
\end{thebibliography}%


\begin{thebibliography}{11}%
\makeatletter
\providecommand \@ifxundefined [1]{%
 \@ifx{#1\undefined}
}%
\providecommand \@ifnum [1]{%
 \ifnum #1\expandafter \@firstoftwo
 \else \expandafter \@secondoftwo
 \fi
}%
\providecommand \@ifx [1]{%
 \ifx #1\expandafter \@firstoftwo
 \else \expandafter \@secondoftwo
 \fi
}%
\providecommand \natexlab [1]{#1}%
\providecommand \enquote  [1]{``#1''}%
\providecommand \bibnamefont  [1]{#1}%
\providecommand \bibfnamefont [1]{#1}%
\providecommand \citenamefont [1]{#1}%
\providecommand \href@noop [0]{\@secondoftwo}%
\providecommand \href [0]{\begingroup \@sanitize@url \@href}%
\providecommand \@href[1]{\@@startlink{#1}\@@href}%
\providecommand \@@href[1]{\endgroup#1\@@endlink}%
\providecommand \@sanitize@url [0]{\catcode `\\12\catcode `\$12\catcode
  `\&12\catcode `\#12\catcode `\^12\catcode `\_12\catcode `\%12\relax}%
\providecommand \@@startlink[1]{}%
\providecommand \@@endlink[0]{}%
\providecommand \url  [0]{\begingroup\@sanitize@url \@url }%
\providecommand \@url [1]{\endgroup\@href {#1}{\urlprefix }}%
\providecommand \urlprefix  [0]{URL }%
\providecommand \Eprint [0]{\href }%
\providecommand \doibase [0]{http://dx.doi.org/}%
\providecommand \selectlanguage [0]{\@gobble}%
\providecommand \bibinfo  [0]{\@secondoftwo}%
\providecommand \bibfield  [0]{\@secondoftwo}%
\providecommand \translation [1]{[#1]}%
\providecommand \BibitemOpen [0]{}%
\providecommand \bibitemStop [0]{}%
\providecommand \bibitemNoStop [0]{.\EOS\space}%
\providecommand \EOS [0]{\spacefactor3000\relax}%
\providecommand \BibitemShut  [1]{\csname bibitem#1\endcsname}%
\let\auto@bib@innerbib\@empty
\bibitem [{\citenamefont {Sieberer}\ \emph {et~al.}(2016)\citenamefont
  {Sieberer}, \citenamefont {Buchhold},\ and\ \citenamefont
  {Diehl}}]{Sieberer_review_sm}%
  \BibitemOpen
  \bibfield  {author} {\bibinfo {author} {\bibfnamefont {L.~M.}\ \bibnamefont
  {Sieberer}}, \bibinfo {author} {\bibfnamefont {M.}~\bibnamefont {Buchhold}},
  \ and\ \bibinfo {author} {\bibfnamefont {S.}~\bibnamefont {Diehl}},\ }\href
  {http://stacks.iop.org/0034-4885/79/i=9/a=096001} {\bibfield  {journal}
  {\bibinfo  {journal} {Rep. Prog. Phys.}\ }\textbf {\bibinfo {volume} {79}},\
  \bibinfo {pages} {096001} (\bibinfo {year} {2016})}\BibitemShut {NoStop}%
\bibitem [{\citenamefont {Kamenev}(2011)}]{Kamenev_book_sm}%
  \BibitemOpen
  \bibfield  {author} {\bibinfo {author} {\bibfnamefont {A.}~\bibnamefont
  {Kamenev}},\ }\href {\doibase 10.1017/CBO9781139003667} {\emph {\bibinfo
  {title} {Field Theory of Non-Equilibrium Systems}}}\ (\bibinfo  {publisher}
  {Cambridge University Press},\ \bibinfo {year} {2011})\BibitemShut {NoStop}%
\bibitem [{\citenamefont {Lebrat}\ \emph {et~al.}(2018)\citenamefont {Lebrat},
  \citenamefont {Gri\ifmmode~\check{s}\else \v{s}\fi{}ins}, \citenamefont
  {Husmann}, \citenamefont {H\"ausler}, \citenamefont {Corman}, \citenamefont
  {Giamarchi}, \citenamefont {Brantut},\ and\ \citenamefont
  {Esslinger}}]{Lebrat2018_sm}%
  \BibitemOpen
  \bibfield  {author} {\bibinfo {author} {\bibfnamefont {M.}~\bibnamefont
  {Lebrat}}, \bibinfo {author} {\bibfnamefont {P.}~\bibnamefont
  {Gri\ifmmode~\check{s}\else \v{s}\fi{}ins}}, \bibinfo {author} {\bibfnamefont
  {D.}~\bibnamefont {Husmann}}, \bibinfo {author} {\bibfnamefont
  {S.}~\bibnamefont {H\"ausler}}, \bibinfo {author} {\bibfnamefont
  {L.}~\bibnamefont {Corman}}, \bibinfo {author} {\bibfnamefont
  {T.}~\bibnamefont {Giamarchi}}, \bibinfo {author} {\bibfnamefont {J.-P.}\
  \bibnamefont {Brantut}}, \ and\ \bibinfo {author} {\bibfnamefont
  {T.}~\bibnamefont {Esslinger}},\ }\href {\doibase 10.1103/PhysRevX.8.011053}
  {\bibfield  {journal} {\bibinfo  {journal} {Phys. Rev. X}\ }\textbf {\bibinfo
  {volume} {8}},\ \bibinfo {pages} {011053} (\bibinfo {year}
  {2018})}\BibitemShut {NoStop}%
\bibitem [{\citenamefont {Giamarchi}(2004)}]{Giamarchi_book_sm}%
  \BibitemOpen
  \bibfield  {author} {\bibinfo {author} {\bibfnamefont {T.}~\bibnamefont
  {Giamarchi}},\ }\href {https://books.google.de/books?id=1MwTDAAAQBAJ} {\emph
  {\bibinfo {title} {Quantum Physics in One Dimension}}}\ (\bibinfo
  {publisher} {Oxford University Press},\ \bibinfo {year} {2004})\BibitemShut
  {NoStop}%
\bibitem [{\citenamefont {Kane}\ and\ \citenamefont
  {Fisher}(1992{\natexlab{a}})}]{Kane1992_sm}%
  \BibitemOpen
  \bibfield  {author} {\bibinfo {author} {\bibfnamefont {C.~L.}\ \bibnamefont
  {Kane}}\ and\ \bibinfo {author} {\bibfnamefont {M.~P.~A.}\ \bibnamefont
  {Fisher}},\ }\href {\doibase 10.1103/PhysRevLett.68.1220} {\bibfield
  {journal} {\bibinfo  {journal} {Phys. Rev. Lett.}\ }\textbf {\bibinfo
  {volume} {68}},\ \bibinfo {pages} {1220} (\bibinfo {year}
  {1992}{\natexlab{a}})}\BibitemShut {NoStop}%
\bibitem [{\citenamefont {Kane}\ and\ \citenamefont
  {Fisher}(1992{\natexlab{b}})}]{Kane1992Long_sm}%
  \BibitemOpen
  \bibfield  {author} {\bibinfo {author} {\bibfnamefont {C.~L.}\ \bibnamefont
  {Kane}}\ and\ \bibinfo {author} {\bibfnamefont {M.~P.~A.}\ \bibnamefont
  {Fisher}},\ }\href {\doibase 10.1103/PhysRevB.46.15233} {\bibfield  {journal}
  {\bibinfo  {journal} {Phys. Rev. B}\ }\textbf {\bibinfo {volume} {46}},\
  \bibinfo {pages} {15233} (\bibinfo {year} {1992}{\natexlab{b}})}\BibitemShut
  {NoStop}%
\bibitem [{\citenamefont {Mitra}\ and\ \citenamefont
  {Giamarchi}(2011)}]{Mitra2011_sm}%
  \BibitemOpen
  \bibfield  {author} {\bibinfo {author} {\bibfnamefont {A.}~\bibnamefont
  {Mitra}}\ and\ \bibinfo {author} {\bibfnamefont {T.}~\bibnamefont
  {Giamarchi}},\ }\href {\doibase 10.1103/PhysRevLett.107.150602} {\bibfield
  {journal} {\bibinfo  {journal} {Phys. Rev. Lett.}\ }\textbf {\bibinfo
  {volume} {107}},\ \bibinfo {pages} {150602} (\bibinfo {year}
  {2011})}\BibitemShut {NoStop}%
\bibitem [{\citenamefont {Schir\'o}\ and\ \citenamefont
  {Mitra}(2015)}]{Schiro2015_sm}%
  \BibitemOpen
  \bibfield  {author} {\bibinfo {author} {\bibfnamefont {M.}~\bibnamefont
  {Schir\'o}}\ and\ \bibinfo {author} {\bibfnamefont {A.}~\bibnamefont
  {Mitra}},\ }\href {\doibase 10.1103/PhysRevB.91.235126} {\bibfield  {journal}
  {\bibinfo  {journal} {Phys. Rev. B}\ }\textbf {\bibinfo {volume} {91}},\
  \bibinfo {pages} {235126} (\bibinfo {year} {2015})}\BibitemShut {NoStop}%
\bibitem [{\citenamefont {Buchhold}\ and\ \citenamefont
  {Diehl}(2015)}]{Buchhold2015_sm}%
  \BibitemOpen
  \bibfield  {author} {\bibinfo {author} {\bibfnamefont {M.}~\bibnamefont
  {Buchhold}}\ and\ \bibinfo {author} {\bibfnamefont {S.}~\bibnamefont
  {Diehl}},\ }\href {\doibase 10.1103/PhysRevA.92.013603} {\bibfield  {journal}
  {\bibinfo  {journal} {Phys. Rev. A}\ }\textbf {\bibinfo {volume} {92}},\
  \bibinfo {pages} {013603} (\bibinfo {year} {2015})}\BibitemShut {NoStop}%
\bibitem [{\citenamefont {Yue}\ \emph {et~al.}(1994)\citenamefont {Yue},
  \citenamefont {Glazman},\ and\ \citenamefont {Matveev}}]{Yue1994_sm}%
  \BibitemOpen
  \bibfield  {author} {\bibinfo {author} {\bibfnamefont {D.}~\bibnamefont
  {Yue}}, \bibinfo {author} {\bibfnamefont {L.~I.}\ \bibnamefont {Glazman}}, \
  and\ \bibinfo {author} {\bibfnamefont {K.~A.}\ \bibnamefont {Matveev}},\
  }\href {\doibase 10.1103/PhysRevB.49.1966} {\bibfield  {journal} {\bibinfo
  {journal} {Phys. Rev. B}\ }\textbf {\bibinfo {volume} {49}},\ \bibinfo
  {pages} {1966} (\bibinfo {year} {1994})}\BibitemShut {NoStop}%
\bibitem [{\citenamefont {Matveev}\ \emph {et~al.}(1993)\citenamefont
  {Matveev}, \citenamefont {Yue},\ and\ \citenamefont
  {Glazman}}]{Matveev1993_sm}%
  \BibitemOpen
  \bibfield  {author} {\bibinfo {author} {\bibfnamefont {K.~A.}\ \bibnamefont
  {Matveev}}, \bibinfo {author} {\bibfnamefont {D.}~\bibnamefont {Yue}}, \ and\
  \bibinfo {author} {\bibfnamefont {L.~I.}\ \bibnamefont {Glazman}},\ }\href
  {\doibase 10.1103/PhysRevLett.71.3351} {\bibfield  {journal} {\bibinfo
  {journal} {Phys. Rev. Lett.}\ }\textbf {\bibinfo {volume} {71}},\ \bibinfo
  {pages} {3351} (\bibinfo {year} {1993})}\BibitemShut {NoStop}%
\end{thebibliography}%

\onecolumngrid
\section{\large Supplemental Material}

\setcounter{equation}{0}
\renewcommand{\theequation}{S\arabic{equation}}
\renewcommand{\bibnumfmt}[1]{[S#1]}
\renewcommand{\citenumfont}[1]{S#1}


\section{I. Exact results for the non-interacting continuum model with localized loss}
\subsection{Retarded Green's function}
\label{sec:retarded}
The quantum master equation (see Eq.(2) in the main text) can be mapped onto a Keldysh action $S = S_0 + S_\text{loss}$~\citesm{Sieberer_review_sm, Kamenev_book_sm}, with
\begin{equation}
S_0 = \int_{x,t}
\left[
i\psi_+^*\dot{\psi}_+ - H(\psi_+^*,\psi_+) - i\psi_-^*\dot{\psi}_- + H(\psi_-^*,\psi_-)
\right],
\end{equation}
and
\begin{equation}
\label{eq:Lindblad-to-Keldysh}
S_\text{loss} = - i  \int_{x,t} \Gamma(x)  \left[L_-^*  L_+ - \frac{1}{2} \left( L_+^* L_+ + L_-^* L_- \right)  \right],
\end{equation}
where $\Gamma (x) = \gamma \delta(x)$ and $L_\pm(x) = \psi_\pm(x)$.
The retarded Green's function $G(x,x',\omega)$ can be then computed conveniently by using the Dyson's equation~\citesm{Kamenev_book_sm}:  
\begin{equation}
\label{eq:Dyson}
G(x,x',\omega) = G_0(x,x',\omega) + \int_y G_0(x,y,\omega)\Sigma(y)G(y,x',\omega) ,
\end{equation}
where 
\begin{equation}
G_0(x,x',\omega) = \frac{1}{2i\sqrt{\omega}} \ee^{i\sqrt{\omega}|x-x'|}
\label{eq:G0xx} 
\end{equation}
is the (translation invariant) retarded Green's function of the system in absence of localized loss, while $\Sigma(y)$ is the self-energy associated with the localized loss. Since the loss appear as a quadratic operator in the Keldysh action, the self-energy is a field-independent function which reads $\Sigma(y) = -i\gamma\delta(y)$, so that Eq.~\eqref{eq:Dyson} can be rewritten as
\begin{equation}
G(x,x',\omega) = G_0(x,x',\omega) -i\gamma G_0(x,0,\omega)G(0,x',\omega), 
\end{equation}
which can readily be solved, yielding
\begin{equation}
\label{eq:Gxx}
G(x,x',\omega)  = G_0(x,x',\omega) -i\gamma\frac{G_0(x,0,\omega)G_0(0,x',\omega)}{1+i\gamma G_0(0,0,\omega)} = \frac{1}{2i\sqrt{\omega}} \left[ \ee^{i\sqrt{\omega}|x-x'|} + r(\omega)\,  \ee^{i\sqrt{\omega}(|x|+|x'|)} \right],
\end{equation}
where 
\begin{equation}
r(\omega) = \frac{-\gamma}{2\sqrt{\omega} + \gamma}
\label{eq:romega} 
\end{equation}
is a function to be discussed below. We report, for later convenience, also the expression for the Green's function in a mixed momentum-real space representation $G(x,q,\omega)$:
\begin{equation}
\label{eq:Gxq}
G(x,q,\omega) = \frac{1}{\omega+i\epsilon - q^2}\left[ \ee^{iqx} + r(\omega)\,  \ee^{i\sqrt{\omega}|x|} \right],
\end{equation}
with $\epsilon$ an infinitesimal dissipation needed in order to ensure causality.

\subsection{Transmission and reflection coefficients}

The transmission and reflection coefficients for the scattering of a particle off the lossy barrier can be inferred from Eqs.~\eqref{eq:Gxx} and~\eqref{eq:Gxq}. For instance, Eq.~\eqref{eq:Gxq} can be interpreted as the response to an external field $h(x,t) = h\exp(iqx-i\omega t)$ corresponding to a plane wave incoming from left. From the on-shell condition $\omega \simeq q^2$ one finds
\begin{equation}
G(x,q,\omega\simeq q^2) \propto 
\begin{cases}
\ee^{iqx} + r(q^2)\ee^{-iqx} & x<0 \\
t(q^2)\ee^{iqx} & x > 0	,
\end{cases} 	
\label{eq:GScattering} 
\end{equation}
where $r(q^2)$ is defined in Eq.~\eqref{eq:romega} and corresponds to the reflection coefficient, while $t(q^2) = 1+r(q^2)$ corresponds to the transmission coefficient. Using Eq.~\eqref{eq:romega} one finds
\begin{equation}
t(q^2) = \frac{2|q|}{2|q| + \gamma}, \quad  r(q^2) = -\frac{\gamma}{2|q| + \gamma}.
\end{equation}

\subsection{One-particle correlation function}
We study the evolution of the one-particle correlation function $C(x,y,t,t') = \langle \psi^\dagger(x,t)\psi(y,t')\rangle$ after the quench of the impurity at time $t_0$. Since the wire is coupled to an empty bath which absorbs irreversibly particles from it, $C(x,y,t,t')$ can be written as a propagation of the  correlation at time $t_0$, by using the retarded Green's function previously computed:
\begin{equation}
\label{eq:Cxy-definition}
C(x,y,t,t') = \int_{x',y'} G^*(x,x',t-t_0)G(y,y',t'-t_0) C(x',y',t_0,t_0).
\end{equation}
With the initial condition $C(x,y,t_0,t_0) = \int_q\ee^{iq(y-x)}n_F(q)$, where $n_F(q)$ is the Fermi distribution at $T=0$, Eq.~\eqref{eq:Cxy-definition} becomes
\begin{equation}
C(x,y,t,t') = \int_{-k_F}^{k_F}\frac{\dd q}{2\pi}\, G^*(x,-q,t)G(y,-q,t').
\end{equation}
We now compute the stationary value of $C(x,y,t,t')$ by taking the limit $t_0\to -\infty$. 
For the sake of clarity, we consider the case $x=y$ and $t=t'$, but the reasoning can be straightforwardly generalized to the case $x\neq y$ and $t\neq t'$. 
\begin{align}
|G(x,-q,t-t_0)|^2 
& = \left| \int\frac{\dd\omega}{2\pi}\, \frac{\ee^{-i\omega(t-t_0)}}{\omega + i\epsilon - q^2}\left[\ee^{-iqx} + r(\omega)\ee^{i\sqrt{\omega}|x|}  \right]   \right|^2 \nonumber \\
& = \left| \int\frac{\dd\omega}{2\pi} \ee^{-i\omega(t-t_0)} \left[\ee^{-iqx} + r(\omega)\ee^{i\sqrt{\omega}|x|}  \right] \int\dd t'\, (-i)\theta(t') \ee^{i(\omega -q^2)t'}\right|^2 \nonumber \\
& = \left|\int\frac{\dd\omega}{2\pi} \ee^{-i\omega t} \left[\ee^{-iqx} + r(\omega)\ee^{i\sqrt{\omega}|x|}  \right]\,2\pi\, \delta(\omega-q^2)\right|^2 \nonumber \\
& = \left| \ee^{-iqx} + r(q^2)\ee^{i|q||x|} \right|^2	,
\label{eq:CalcG} 
\end{align}
were in the third step we took the limit $t_0\to-\infty$. The local density of particles at time $t$ is given by $n(x,t) = C(x,x,t,t)$. By making use of the previous equation we obtain 
\begin{equation}
n(x) = \frac{1}{\pi}\int_0^{k_F}\dd q\bigg[ 1+r(q^2)^2 + r(q^2) + r(q^2) \cos(2q|x|) \bigg] \equiv n_\text{ness} + \delta n(x), \label{eq:densityProfile} 
\end{equation}
where $n_\text{ness}$ is the homogeneous background and $\delta n(x)$ the density modulations, which exhibits for $|x|\gg k_F^{-1}$ Friedel oscillations
\begin{equation}
\delta n(x) = r(k_F) \frac{\sin(2k_F|x|)}{2\pi|x|}.
\end{equation}

\subsection{Preservation of the Fermi momentum and experimental observability}
\begin{figure}[ht]
\centering
\includegraphics[width=9cm]{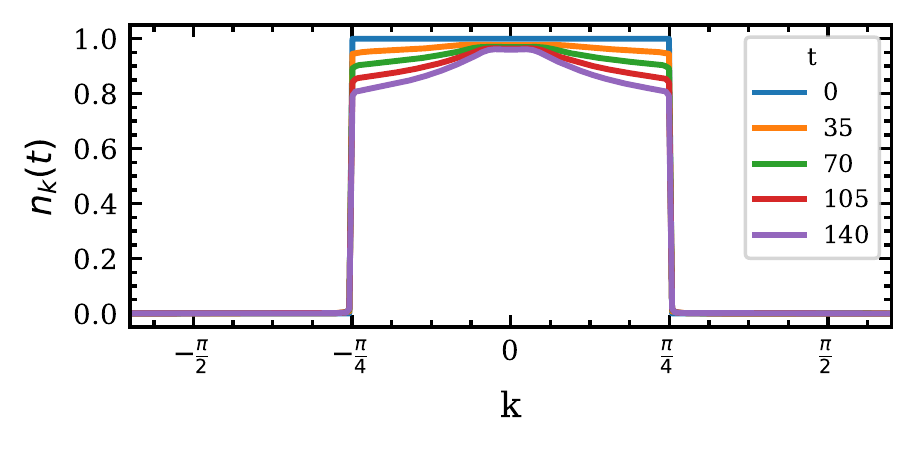}
\caption{
(Color online). Numerical study of the momentum occupation in the second time regime for different times after the quench ($L=401, \gamma = 1, N(0)/L=0.25$ ).
}
\label{fig:fig1SM}
\end{figure}

We investigate the time-evolution of the momentum occupation number in the non-interacting system after a quench of the localized loss both in the lattice system and in the continuum model. 
In the lattice system we evaluate $n_k(t) = \langle \psi_k^\dagger \psi_k \rangle$ numerically. The system is initialized in its zero-temperature ground state characterized by a Fermi momentum $k_F=\pi N(0)/L$, with $N(0)$ the initial particle number and $L$ the system size. Crucially, we observe in the dynamics after the quench that the step at the initial Fermi momentum $k_F$ remains well-defined (see Fig.~\ref{fig:fig1SM}), although the overall number of particles depletes. At all times the distribution exhibits a discontinuity at the initial Fermi momentum.

In order to obtain an analytical expression for the momentum distribution $n(k,t)$, we evaluate $ \langle \psi^\dagger(k',t) \psi(k,t) \rangle$ for the continuum model. By taking the Fourier transform of Eq.~\eqref{eq:Cxy-definition} we obtain:
\begin{equation}
 \langle \psi^\dagger(k,t) \psi(k,t) \rangle = \int \frac{\dd q}{2\pi} G^*(k,-q,t-t_0) G(k,-q,t-t_0) n_0(q)	 ,
\end{equation}
with $ n_0(q) = \theta(k_F^2-q^2)$ the initial zero-temperature Fermi distribution.
We are left to evaluate
\begin{multline}
\label{eq:momentum-distro}
\langle \psi^\dagger(k,t) \psi(k,t) \rangle \\
= \int \frac{\dd q}{2\pi} n_0(q) \left[ |G_0(k,t-t_0)|^2 \delta^2(k+q) + \delta(k+q) (G_0^*(k,t-t_0) \widetilde G(k,-q,t-t_0) +c.c.) + |\widetilde G(k,-q,t-t_0)|^2 \right],
\end{multline}
with
\begin{equation}
G(k,k',\omega) = \delta(k-k') G_0 (k,\omega)  + \widetilde G(k,k',\omega) ,\qquad \widetilde G(k,k',\omega) =  i2 \sqrt{\omega} r(\omega) G_0(k ,\omega) G_0(k' ,\omega)	,
\end{equation}
where $G_0(k,\omega)$ is the Fourier transform of Eq.~\eqref{eq:G0xx}, and $r(\omega)$ is given in Eq.~\eqref{eq:romega}.
By noticing that $n(k,t)$ and $\langle \psi^\dagger(k,t) \psi(k,t) \rangle$ are proportional by a factor diverging as the system's volume (hence the square delta function in Eq.~\eqref{eq:momentum-distro}), and by taking the limit $t_0 \rightarrow -\infty$, we obtain the stationary distribution
\begin{equation}
n(k) = n_0(k) \left( 1 - \frac{\gamma / k}{ 1 + \gamma / (2k)} + \frac{1}{2} \frac{\gamma^2 / k^2}{(1+ \gamma/(2k) )^2}  \right) = n_0(k) (1- \eta_0(k)),
\label{eq:MomentumDistEta} 
\end{equation}
where we recognized the escape probability $\eta_0(k)$. Hence, we find that the discontinuity at the initial Fermi momentum $k_F$ persists in the stationary state, as $\eta_0(k)$ is a smooth function of $k$.
Eq.~\eqref{eq:MomentumDistEta} can be generalized to account for interactions.

In experiments where the stationary regime could approximately be reached one could obtain the value of $\eta(k)$ by measuring the momentum distribution of the system in the presence of the
impurity, and by comparing it to the one without impurity.
An analogous situation to the discussed stationary regime could be achieved in systems coupled to reservoirs at the ends (cf. Ref.~\citesm{Lebrat2018_sm}), both at $T=0$ and $\mu = k_F$, thus sustaining the particle currents.
%
%

\section{II. Bosonization of a dissipative impurity}
%
%
The low-energy properties of a system interacting one-dimensional fermions is captured by the Luttinger Hamiltonian described in the main text (see Eq.(3)). The mapping between fermions and bosons is done via the transformation (considering the two leading harmonics)~\citesm{Giamarchi_book_sm, Kane1992_sm, Kane1992Long_sm}: 
\begin{equation}
\label{eq:bosonization-mapping}
\psi \sim \sqrt{\rho_0} \, \left[ e^{i k_F x}  e^{i (\phi + \theta)} +  e^{-i k_F x} e^{i (\phi - \theta)}	\right],
\end{equation}
where $\phi$ and $\theta$ are bosonic fields describing phase fluctuations and density fluctuations, respectively. The associated Keldysh action $S_0$ can be written as~\citesm{Mitra2011_sm,Schiro2015_sm, Buchhold2015_sm}
\begin{equation}
S_0 = \int_{k,\omega}\ \chi^\dagger (k, \omega) G^{-1}(k, \omega) \chi(k, \omega)
\end{equation}
where $\int_{k,\omega} \equiv \int \dd k \dd \omega/(2\pi)^2 $,  $\chi \equiv (\phi_c, \theta_c,  \phi_q,  \theta_q )^T$ ,
\begin{equation}
 G^{-1}(k, \omega) = 
 \begin{pmatrix}
0 & P_A	(k, \omega)\\
 P_R(k, \omega) &  P_K(k, \omega)
\end{pmatrix},	
\end{equation}
and
\begin{equation}
\label{eq:PR-PK}
P_R = P_A^\dagger = \frac{1}{2\pi} 
 \begin{pmatrix}
 v g k^2 + i \epsilon \omega & - k \omega \\
 - k \omega &  v g^{-1} k^2 + i \epsilon \omega
 \end{pmatrix}	
 ,\qquad
P_K =  \frac{1}{2\pi}
 \begin{pmatrix}
 2 i \epsilon \omega \coth \frac{\omega}{2 T} & 0 \\
 0 &  2 i \epsilon \omega \coth \frac{\omega}{2 T}
 \end{pmatrix}. 
\end{equation}
The retarded and Keldysh Green's functions, $G_R$ and $G_K$, respectively, can be obtained as  $G_R = P_R^{-1}$ and $G_K = P_K^{-1}$~\citesm{Kamenev_book_sm}.
By using the mapping~\eqref{eq:Lindblad-to-Keldysh} and the bosonization formula~\eqref{eq:bosonization-mapping}, the  backward scattering due to the impurity reads:
\begin{equation}
\label{eq:SD}
S_\text{back} =  - i  2 \gamma  \int_{x,t} \delta(x) \left[ \cos \theta_c  \left( \ee^{i\phi_q} - \cos  \theta_q  \right)  \right], 
\end{equation} 
where $\theta_{c,q} = \theta_+ \pm \theta_-$ and $\phi_{c,q} = \phi_+ \pm \phi_-$.

A renormalization group analysis is then performed by integrating out fast modes lying in the momentum-shells $k \in \left[ \Lambda \ee^{-\ell}, \Lambda \right]$ and subsequently rescaling as $(x,t) \rightarrow (\ee^\ell x, \ee^\ell t)$. In the weak coupling limit ($\gamma \to 0 $) we consider the leading term in the expansion in $\gamma$: 
\begin{equation}
\langle \ee^{  i S_\text{back} } \rangle_\text{fast}  \simeq \langle 1 +  i S_\text{back} \rangle_\text{fast} \simeq  \ee^{  \langle i S_\text{back}\rangle_\text{fast}  }	.
\end{equation}
By making use of the Gaussian identity $\langle \ee^{i x} \rangle = \exp[ - \langle x^2\rangle/2 ]$ and the correlation functions from Eq.~\eqref{eq:PR-PK}, we obtain from Eq.~\eqref{eq:SD}
\begin{equation}
\left\langle   S_\text{loss}[\phi_\text{fast} + \phi_\text{slow} , \theta_\text{fast} + \theta_\text{slow}]   \right\rangle_\text{fast}  =  
  S_\text{loss}[\phi_\text{slow}, \theta_\text{slow} ] \ee^{- \left\langle \theta_{c}^2(x,t) \right\rangle_\text{fast} }	.
\end{equation}
This first order corrections can thus be calculated from~\eqref{eq:PR-PK}, and for $T=0$ it reads
\begin{equation}
\langle \theta_c^2(x,t)\rangle_\text{fast}  =  g \int_{\Lambda \ee^{-\ell}}^\Lambda  \frac{\dd k}{k} = g \ell	.
\end{equation}
Finally, by noticing that the canonical scaling dimension of the dissipation strength $\gamma$ is $[\gamma] = 1$, we obtain the flow equation (5) in the main text.

\section{III. Renormalization of scattering amplitudes}
We proceed by computing the corrections of the scattering probabilities $\mathcal{T}$ and $\mathcal{R}$ due to interactions $V$ (see Eq. (1) in the main text) to first order in perturbation theory in the interactions. To this end, its convenient to focus on the transmission and reflection amplitudes, which relate to the corresponding probabilities via $\mathcal{T} = |t|^2$ and $\mathcal{R} = |r|^2$. $t$ and $r$ are defined via the retarded Green's function as discussed in Sec.~I. 
The corrections can be obtained from the perturbed retarded Green's function $G = G_{0} +  \delta G$, where $G_0$ denotes the unperturbed Green's function ($V=0$): 
\begin{equation}
\delta G(x,y,\omega) = \int_{x',y'} G_0(x,x',\omega)\left[ V_H(x',y') + V_{ex}(x',y') \right] G_0(y',y,\omega)		,
\label{eq:GPert} 
\end{equation}
where the Hartree and exchange potentials $V_H$ and $V_{ex}$ are given by
\begin{equation}
V_H(x,y) = \delta(x-y)\int_{x'}V(x-x')C(x',x',t,t), \qquad V_{ex}(x,y) = -V(x-y)\,C(y,x,t,t) ,
\end{equation}
with $C(x,y,t,t)$ evaluated in the stationary limit. From Eq.~\eqref{eq:GPert} one obtains the following corrections (see Ref.~\citesm{Yue1994_sm}),
\begin{subequations}
\begin{align}
t & = t_0 + \alpha\, t_0r_0^2 \, \log|d(k-k_F)| ,\\
r & = r_0 + \frac{\alpha}{2} r_0\bigg(r_0^2 + t_0^2 - 1 \bigg) \, \log|d(k-k_F)|,
\end{align}
\end{subequations}
with  $t,r \equiv t(k), r(k)$ for  $k\sim k_F$, $d$ is the typical length scale of $V(x)$, and $\alpha = [V(2k_F)-V(0)]/(2\pi v_F)$. 
The logarithmic divergences of the perturbation theory remain at higher orders and are cured by a proper resummation, achieved by a real space or a frequency RG \citesm{Yue1994_sm, Matveev1993_sm} leading to
\begin{subequations}
\label{eq:RG-equationsSM}
\begin{align}
\frac{\dd t}{\dd \ell} & = - \alpha \, t r^2	, \\
\frac{\dd r}{\dd \ell} & = - \frac{\alpha}{2} \,  r \bigg( t^2+r^2 - 1 \bigg)	.
\end{align}
\end{subequations}
From here it is straightforward to derive the flow equations of the scattering probabilities (Eqs.~(8) in the main text).

Since both, the continuity relation $t=1+r$ is preserved and $t,r$ remain real-valued along the flow, it is possible to parametrize the flow of $r$ and $t$ by a single function $\widetilde{\gamma}$, such that
\begin{equation}
r = -\frac{\widetilde{\gamma}}{1+\widetilde{\gamma}}, \qquad t = \frac{1}{1+\widetilde{\gamma}}.
\end{equation}
The flow equation for $\widetilde{\gamma}$ is then easily derived from Eq.~\eqref{eq:RG-equationsSM} and reads
\begin{equation}
\frac{\dd \widetilde{\gamma}}{\dd \ell} = \alpha \frac{\widetilde{\gamma}^2}{1+\widetilde{\gamma}}.
\end{equation}

\section{IV. Input-output formalism}

We investigate the loss of particles from the system at $x=0$. To this end, we study $G(x=0, p, \omega)$,  describing the response at $x=0$ to a plane wave perturbation, as discussed in Sec.~I (cf. Eq.~\eqref{eq:Gxq}):
\begin{equation}
\label{eq:t_eta}
G(x=0, p, \omega) = t_{\eta}(\omega) G_0(p, \omega) 		.
\end{equation}
Here, we introduced the amplitude $ t_{\eta}$, related to $\eta$ through 
\begin{equation}
\frac{\gamma}{k} t_\eta^2 = 1- \mathcal{T}-\mathcal{R} \equiv \eta.
\label{eq:EtaTmSqIdentity} 
\end{equation}
Eqs.~\eqref{eq:t_eta} and~\eqref{eq:EtaTmSqIdentity} can then be regarded as an operative definition to evaluate $\eta$ directly from the response function.\\
We then introduce $g(t_2 - t_1) = \langle \psi^\dagger(x=0, t_1) \psi(x=0, t_2) \rangle$ 
and obtain as a generalization of the calculation in Sec.~I
\begin{align}
g(t_2 - t_1) &= \int_{-k_F}^{k_F} \frac{dp}{2 \pi}  G(x=0, -p, t_1- t_0) G(x=0, -p, t_2 - t_0) = \int_{-k_F}^{k_F} \frac{dp}{2 \pi} \ee^{- p^2 (t_2 - t_1)} | t_{\eta}(p^2)|^2	.
\end{align}
As discussed in the main text, we consider the wire being coupled to a continuum of free fermionic modes, described by
$H_\text{int} = \sum_{\qq}(g_\qq c^\dagger_\qq \psi_0+ \text{h.c.})$. The momentum resolved loss rate is then related to $t_{\eta}$ by 
\begin{equation}
\frac{\dd \langle c^\dagger_\qq c_\qq\rangle}{\dd t} = |g_\qq|^2  g(\omega=\epsilon_\qq)  = |g_\qq|^2  \frac{\theta(E_F - \epsilon_\qq) }{\sqrt{\epsilon_\qq}} | t_{\eta}(\omega=\epsilon_\qq)|^2	,
\label{eq:momResolvedLossRate} 
\end{equation}
with $E_F = k_F^2/2m$.\\
The renormalization procedure from Sec.~III can be adapted to obtain the renormalization of $t_{\eta}$. As the starting point, the perturbative correction of $G(x=0,p,\omega)$ is given by
\begin{equation}
\delta G(x=0,p,\omega) = \int_{x',y'} G(x=0,x',\omega)\left[ V_H(x',y') + V_{ex}(x',y') \right] G(y',p,\omega).
\end{equation}
Crucially, we observe that $ t_{\eta}$ enters the expression via $G(x=0,x',\omega)$, while the logarithmic divergencies are generated by the asymptotic forms of $G(y',y,\omega)$, $V_H(x',y')$ and $ V_{ex}(x',y')$ in which accordingly $t$ and $r$ appear. \\
The flow equation of $t_\eta \equiv  t_\eta(k)$ for $ k\sim k_F$  is found based on these considerations as
\begin{equation}
\frac{\dd t_\eta}{\dd \ell} = - \frac{\alpha}{2} t_\eta \left( r^2 + t r  \right)		.
\end{equation}
One can readily check that the flow equations of $\eta = 1 - t^2 - r^2$ and $t_\eta^2$ coincide, in agreement with Eq.~\eqref{eq:EtaTmSqIdentity}. 
According to Eqs.~\eqref{eq:momResolvedLossRate} and~\eqref{eq:EtaTmSqIdentity}, we can relate the momentum resolved loss rate to the escape probability $\eta$ as the key loss indicator of the system and arrive at the central result given in Eq.~(11) of the main text.


\bibliographystylesm{apsrev4-1} 
\bibliographysm{bibliosm}

\end{document}